# An Artificial-Noise-Aided Secure Scheme for Hybrid Parallel PLC/Wireless OFDM Systems

Ahmed El Shafie[†], Mohamed F. Marzban[†], Rakan Chabaan[⋆], Naofal Al-Dhahir[†]
[†]University of Texas at Dallas, USA.
[⋆]Hyundai-Kia America Tech. Center, Inc.

*Abstract*—We investigate the physical-layer security of indoor hybrid parallel power-line/wireless orthogonal-frequency division-multiplexing (OFDM) communication systems. We propose an artificial-noise (AN) aided scheme to enhance the system's security in the presence of an eavesdropper by exploiting the decoupled nature of the power-line and wireless communication media. The proposed scheme does not require the instantaneous channel state information of the eavesdropper's links to be known at the legitimate nodes. In our proposed scheme, the legitimate transmitter (Alice) and the legitimate receiver (Bob) cooperate to secure the hybrid system where an AN signal is shared from Bob to Alice on the link with the lower channel-to-noise ratio (CNR) while the information stream in addition to a noisy-amplified version of the received AN signal is transmitted from Alice to Bob on the link with higher CNR at each OFDM sub-channel. In addition, we investigate the effect of the transmit power levels at both Alice and Bob and the power allocation ratio between the data and AN signals at Alice on the secure throughput. We investigate both single-link eavesdropping attacks, where only one link is exposed to eavesdropping attacks, and two-link eavesdropping attacks, where the two links are exposed to eavesdropping attacks.

## I. INTRODUCTION

Hybrid parallel information transmission over two different media has received increased attention recently. For example, the availability of both power-line and wireless links in Smart Grid (SG) networks can be exploited to enhance the data rate and/or reliability of two-way communications [1]–[3].

One of the most critical vulnerabilities in SGs is the confidentiality of the measurement and management information messages [4]. The privacy of the consumers' electricity consumption information can be compromised by intercepting the signal exchanges between consumers and the utility. One application that motivated this work is securing two-way vehicle-to-grid (V2G) communications between electric vehicle charging stations and data aggregators over power-line and unlicensed wireless links [5].

In the seminal work of Wyner [6], a system's physical layer (PHY) security was characterized in terms of the secrecy capacity which was defined as the maximum transmission rate of the legitimate link without information leakage to an eavesdropper. Recently, the authors of [7] showed that the secrecy capacity in power-line communications (PLC) is fairly low compared to wireless communications, especially in the high signal-to-noise ratio (SNR) regime. Motivated by the data rate gains of multiple-input multiple-output (MIMO) in PLC, the authors of [8] showed that MIMO techniques enhance the PHY security of PLC. In [9], the authors investigated the PHY security in multi-carrier and multi-user broadcast systems. The

This work is supported by Hyundai Inc.

effect of the channel statistics on the achievable secrecy rate was analyzed.

Unlike the above-mentioned work, we do not assume that the eavesdropper's channel state information (CSI) is known at the legitimate nodes. Instead, we design an artificial-noise (AN) scheme to enhance the hybrid system's security. Although several recent papers [2], [3] investigated hybrid parallel PLC/wireless systems to enhance reliability and data rates, their PHY security has not been investigated.

The singular-value decomposition (SVD)-based AN-aided scheme was proposed in the wireless PHY-security literature, see [10] and references therein, where AN vectors are transmitted in a direction orthogonal to the data vectors. Since the PLC and wireless media are decoupled, the SVD-based scheme fails to enhance the security because the eavesdropper can always intercept the transmitted data over each sub-channel. This motivates us to propose a new AN-aided scheme that exploits the decoupled nature of hybrid parallel systems. Although we consider a hybrid parallel PLC/wireless system as a case study, our proposed scheme is applicable to other media diversity transmissions with two decoupled communications media.

The contributions of this paper are summarized as follows

- We propose a new AN-aided scheme to improve the PHY security of hybrid parallel PLC/wireless transmissions which is based on the cooperation between the legitimate transmitter (Alice) and legitimate receiver (Bob). We investigate single-link eavesdropping attacks, where only one link is exposed to eavesdropping attacks, and two-link eavesdropping attacks, where the two links are exposed to eavesdropping attacks. We compute the instantaneous secrecy rate (ISR) of the proposed scheme which is then used to evaluate the secure throughput of the hybrid PLC/wireless communication system.
- For the case of two-link eavesdropping, unlike the no-AN scenario, where we show that the ISR depends on the links' CSI and saturates at high Alice's input SNR levels, the ISR of our proposed scheme, at infinite Alice's transmit power, is independent of the eavesdropper's CSI. Hence, the ISR is known at Alice in each channel realization and Alice can adjust the secrecy transmission rate to be equal to the ISR. Moreover, our proposed scheme can achieve any target secrecy rate based on the proper selection of the transmit powers at both Alice and Bob and the power allocation ratio between the data and AN signals at Alice.

## II. SYSTEM MODEL

We consider an indoor hybrid parallel PLC/wireless system as in, e.g., [2], [3], and shown in Fig. 1. A legitimate source node (Alice) communicates with her legitimate destination node (Bob) in the presence of a passive eavesdropping node (Eve). For notational convenience, we will use the subscripts A, B, and E to denote Alice, Bob, and Eve, respectively. For Eve, we assume single-link eavesdropping, where one Eve is located at one medium only (either PLC or wireless), and the two-link eavesdropping, where one eavesdropper is located at each medium. Each node has one wireless transceiver and one PLC transceiver. Alice transmits her data using orthogonal frequency-division multiplexing (OFDM) as will be explained in details in the next section.

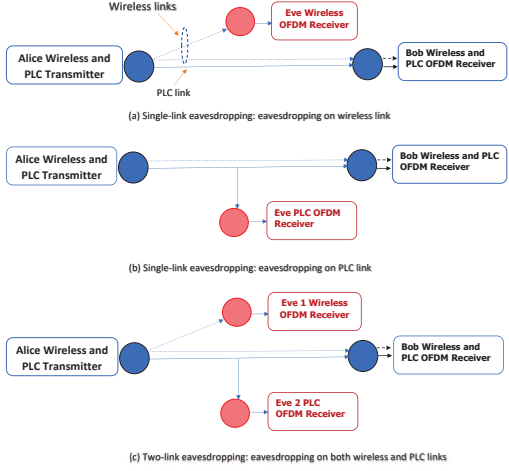

Fig. 1: Hybrid parallel wireless/PLC system.

Alice and Bob perform the channel estimation by sharing a known sequence of pilots at the beginning of each channel coherence interval. Subsequently, Alice and Bob can estimate their channel coefficients while their channel remains unknown at Eve. In addition, we assume that the channel coefficients between Alice and Eve are only known to Eve. Hence, we do not assume the availability of Eve's instantaneous CSI at the legitimate nodes.

In PLC, the links between Alice, Bob and Eve are interdependent. To capture these dependencies, we apply the bottom-up approach as in [8] to compute the channel frequency responses of the PLC links. Although the PLC link is often affected by non-Gaussian additive noise in the time domain, the colored-Gaussian additive noise model is a good model in the frequency domain due to the fast Fourier transform (FFT) processing of the received samples. For the wireless channel, each coefficient of the channel impulse response (CIR) is modeled as a zero-mean circularly-symmetric Gaussian random variable. The channel impulse response (CIR) coefficients of a wireless link are assumed to be independent and identically distributed (i.i.d.) zero-mean circularly-symmetric Gaussian random variables with variance $\sigma^2_{i,m_1-m_2}$ for the $i$th coefficient of the $m_1-m_2$ link, where $m_1, m_2 \in \{A, B, E\}$ and $i = \{0, 1, 2, \ldots, \nu^W_{m_1-m_2}\}$ with $\nu^W_{m_1-m_2}$ denoting the delay spread of the wireless link $m_1-m_2$. The additive noise in the wireless medium is modeled as an additive Gaussian noise (AGN) random process.

## III. PHY SECURITY OF THE HYBRID PARALLEL PLC/WIRELESS SYSTEMS WITHOUT AN INJECTION

Let $\ell \in \{P, W\}$ denote the PLC and wireless media, respectively. Denote the channel coefficient at sub-channel $k$ of the $m_1-m_2$ link over medium $\ell \in \{P, W\}$ by $H^\ell_{m_1-m_2,k}$, where $m_1, m_2 \in \{A, B, E\}$. Alice transmits her data over $N$ orthogonal sub-channels using OFDM. We assume a single data stream transmission over each sub-channel. Since the PLC and wireless media are decoupled, the rate of each sub-channel of the Alice-Bob link is maximized when the Alice-Bob link with the higher channel-to-noise ratio (CNR) over the two media is selected. Hence, the data stream at each sub-channel is transmitted over the link with the higher CNR. We call this scheme '*transmit-selection combining (TSC)*'. Since Alice does not have any knowledge about Eve's instantaneous CSI, she can only optimize her transmit power levels across the OFDM sub-channels to maximize the rate of her own link (i.e. Alice-Bob link). Hence, Alice first computes, at each sub-channel, the maximum of the Alice-Bob link channel gains over the two media. Then, she performs the water-filling algorithm to maximize her transmission rate. For simplicity, we assume equal power allocation over the active sub-channels whose indices are determined from the water-filling algorithm [11]. Assuming that Alice's total transmit power is constrained by $P_A$ Watts, then the power allocated to the $k$th OFDM sub-channel, denoted by $p_{A,k}$, is

$$p_{A,k} = \begin{cases} P_A/|\mathcal{A}|, & \text{if the } k\text{th OFDM sub-channel is active} \\ 0, & \text{otherwise} \end{cases} \quad (1)$$

where $\mathcal{A}$ is the set of active sub-channels, $|\cdot|$ denotes either absolute value or set cardinality depending on the context in which it is used, and $|\mathcal{A}|$ is the number of active sub-channels ($|\mathcal{A}| \leq N$). We emphasize that optimizing the power levels over the active OFDM sub-channels to maximize the rate of the Alice-Bob link will add computational complexity to the problem due to the iterative solution without a significant rate gain over our assumed equal-power allocation over the active sub-channels, as was shown in [11]. For single-link eavesdropping, the achievable rates at both Bob and Eve over sub-channel $k$ are thus given by

$$\begin{aligned} R_{A-B,k} &= \log_2\left(1 + |H^{i^k_H}_{A-B,k}|^2 \Gamma^{i^k_H}_{A-B,k}\right), \ k \in \mathcal{A} \\ R_{A-E,k} &= \mathbb{I}_{i^k_H} \log_2\left(1 + |H^{i^k_H}_{A-E,k}|^2 \Gamma^{i^k_H}_{A-E,k}\right), \ k \in \mathcal{A} \end{aligned} \quad (2)$$

where $i^k_H \in \{P, W\}$ ($i^k_L \in \{P, W\}$) denotes the medium where the Alice-Bob link has the higher (lower) CNR over sub-channel $k$, $\Gamma^\ell_{m_1-m_2,k}$ is the input SNR at OFDM sub-channel $k$ over medium $\ell \in \{P, W\}$ of the $m_1-m_2$ link while $\mathbb{I}_{i^k_H}$ is given by

$$\mathbb{I}_{i^k_H} = \begin{cases} 1, & \text{if Eve eavesdrops on link } i^k_H \text{ (i.e. } i_E = i^k_H \text{ )} \\ 0, & \text{otherwise} \end{cases} \quad (3)$$

where $i_E \in \{P, W\}$ is the link over which Eve is eavesdropping. For the quasi-static channel model assumed in this paper, and since Eve's CSI is not known at Alice, Alice transmits her data with a target/fixed secrecy rate $\mathcal{R}$ bits per OFDM block. If $\mathcal{R}$ is less than or equal to the rate of the Alice-Bob link, Alice transmits her data to Bob. Otherwise, she remains silent during the current transmission block to avoid transmission outage. If $\mathcal{R}$ is higher than the ISR, then there is secrecy outage and the secrecy is compromised. If $\mathcal{R}$ is less than or equal to the ISR, then there is no secrecy outage and the data is decoded at Bob and secured from Eve. The system's security performance is measured by the secure throughput, which is defined to be the number of securely decoded bits at Bob per second per Hertz. First, we compute the ISR, then we use it to calculate the secure throughput.

For an OFDM-based system, the ISR is given by

$$R_{\text{A}-\text{B}}^{\text{sec}} = \left[ \sum_{k \in \mathcal{A}} R_{\text{A}-\text{B},k} - \sum_{k \in \mathcal{A}} R_{\text{A}-\text{E},k} \right]^+ \quad (4)$$

where $[\cdot]^+ = \max\{\cdot, 0\}$. The ISR of the hybrid parallel PLC/wireless system under single-link eavesdropping is thus given by

$$R_{\text{A}-\text{B}}^{\text{sec}} = \left[ \sum_{k \in \mathcal{A}} \log_2 \left( \frac{1 + |H_{\text{A}-\text{B},k}^{i_H^k}|^2 \Gamma_{\text{A}-\text{B},k}^{i_H^k}}{1 + \mathbb{I}_{i_H^k} |H_{\text{A}-\text{E},k}^{i_H^k}|^2 \Gamma_{\text{A}-\text{E},k}^{i_H^k}} \right) \right]^+ \quad (5)$$

while under two-link eavesdropping is given by

$$R_{\text{A}-\text{B}}^{\text{sec}} = \left[ \sum_{k \in \mathcal{A}} \log_2 \left( \frac{1 + |H_{\text{A}-\text{B},k}^{i_H^k}|^2 \Gamma_{\text{A}-\text{B},k}^{i_H^k}}{1 + |H_{\text{A}-\text{E},k}^{i_H^k}|^2 \Gamma_{\text{A}-\text{E},k}^{i_H^k}} \right) \right]^+ \quad (6)$$

which is finite at infinite input SNR and depends on the instantaneous CSI of the links. At infinite input SNR levels, the secrecy rate of the hybrid parallel PLC/wireless system in (6) is given by

$$R_{\text{A}-\text{B}}^{\text{sec}} = \left[ \sum_{k \in \mathcal{A}} \log_2 \left( \frac{|H_{\text{A}-\text{B},k}^{i_H^k}|^2 \kappa_{\text{E},k}^{i_H^k}}{|H_{\text{A}-\text{E},k}^{i_H^k}|^2 \kappa_{\text{B},k}^{i_H^k}} \right) \right]^+ \quad (7)$$

The secure throughput is defined as follows

$$\mu = \mathcal{R} \Pr\{R_{\text{A}-\text{B}}^{\text{sec}} \geq \mathcal{R}\} \quad (8)$$

where $\Pr\{R_{\text{A}-\text{B}}^{\text{sec}} < \mathcal{R}\}$ denotes the secrecy rate outage probability.

We consider three scenarios for eavesdropping attacks: i) single-link eavesdropping attacks which includes two cases: 1) Eve is eavesdropping on the PLC link only; or 2) Eve is eavesdropping on the wireless link only; and ii) two-link non-colluding eavesdropping attacks where one Eve is eavesdropping on the PLC link and the other Eve is eavesdropping on the wireless link. Due to space limitations, the case of colluding eavesdropping where the eavesdroppers share the received signals is left for future work since its analysis is lengthy.

## IV. PROPOSED AN-AIDED TRANSMISSION SCHEME

In this section, we propose an AN-aided scheme which increases the secure throughput by exploiting the decoupled nature of the hybrid media. In our proposed scheme, which we refer to as the 'AN-sharing scheme', we assume that Bob transmits AN symbols to Alice over each active OFDM sub-channel. At the same time, based on the location of Eve, we have two scenarios. If Eve is located at the medium which has the higher CNR, Bob transmits the AN symbol over the link with the lower CNR, which was not utilized in the no-AN scenario. Otherwise, Bob remains silent since the data is not exposed to eavesdropping and there is no AN injection. Nevertheless, in case of two-link eavesdropping as will be discussed shortly, Bob will always share AN over the lower CNR link to confuse both eavesdroppers. To allow simultaneous reception and transmission over different media (i.e. PLC and wireless) per a sub-channel, and since Alice needs to forward Bob's AN signals, at the beginning of each coherence time, Bob transmits one OFDM block and Alice remains silent. In the following transmission times, Bob sends new AN signals and Alice forwards the previously-received AN symbols and transmits her data. The AN symbols are generated from a random codebook which is only known at Bob. To maximize the rate of the Alice-Bob link, Alice transmits the data symbols over the Alice-Bob link with the higher CNR at each OFDM sub-channel along with an amplified noisy version of the AN symbol received from Bob. Since Bob is the only node who knows the AN codebook and symbols, he can remove the interference caused by the AN symbols prior to information decoding while Eve does not know the used AN codebook and, hence, cannot remove them. The received signal at Alice over sub-channel $k$ is given by

$$\mathcal{A}_k = H_{\text{A}-\text{B},k}^{i_L^k} Z_{\text{B},k} + N_{\text{A},k}^{i_L^k} \quad (9)$$

where $H_{\text{A}-\text{B},k}^{i_L^k}$ is the channel coefficient at sub-channel $k$ of the Alice-Bob link where the (PLC or wireless) medium with the lower CNR is selected. In addition, $Z_{\text{B},k}$ with power $\mathbb{E}\{Z_{\text{B},k} Z_{\text{B},k}^*\} = p_{\text{A},k}^B = \frac{P_\text{B}}{|\mathcal{A}|}$ is the zero-mean AN symbol transmitted from Bob to Alice over sub-channel $k$ with $P_\text{B}$ denoting the total power budget at Bob in Watts, and $N_{\text{A},k}^{i_H^k}$ is the AGN signal at Alice over sub-channel $k$ of medium $i_H^k$.

Alice transmits the following signal over sub-channel $k$

$$\mathcal{T}_k = X_k + \omega_k \mathcal{A}_k = X_k + \omega_k \left( H_{\text{A}-\text{B},k}^{i_L^k} Z_{\text{B},k} + N_{\text{A},k}^{i_L^k} \right) \quad (10)$$

where $X_k$ is the data symbol transmitted over sub-channel $k$ with power $\mathbb{E}\{X_k X_k^*\} = \theta P_\text{A} / |\mathcal{A}|$ where $0 \leq \theta P_\text{A} \leq P_\text{A}$ is the portion of the total power budget $P_\text{A}$ assigned to data transmissions, and $\omega_k$ is the weight coefficient used at sub-channel $k$ for AN transmission. As in the amplify-and-forward relaying scheme, Alice first normalizes the power of the received signal from Bob at sub-channel $k$ using a weight of $1/\sqrt{|H_{\text{A}-\text{B},k}^{i_L^k}|^2 \frac{P_\text{B}}{|\mathcal{A}|} + \kappa_{\text{B},k}^{i_L^k}}$ with $\kappa_{m_2,k}^\ell$ denoting the additive noise power at OFDM sub-channel $k$ over medium $\ell$ of node $m_2 \in \{\text{A}, \text{B}, \text{E}\}$. Then, she multiplies this signal by the AN allocated power coefficient given by $\sqrt{(1-\theta) \frac{P_\text{A}}{|\mathcal{A}|}}$. Thus, $\omega_k = \sqrt{p_{\text{A},k} / \kappa_A^{i_H^k}} \times \sqrt{(1-\theta)} / \sqrt{|H_{\text{A}-\text{B},k}^{i_L^k}|^2 \Gamma_{\text{B}-\text{A},k}^{i_L^k} + 1}$.

The received signal at Bob over OFDM sub-channel $k$ of medium $i_H^k$ is given by

$$Y_{B,k}^{i_H^k} = H_{A-B,k}^{i_H^k} \left(X_k + \omega_k \left(H_{A-B,k}^{i_L^k} Z_{B,k} + N_{A,k}^{i_L^k}\right)\right) + N_{B,k}^{i_H^k} \quad (11)$$

where $N_{B,k}^{i_H^k}$ is the corresponding AGN sample at Bob. Since Bob knows the used AN symbols, he cancels them out prior to data decoding to get the signal

$$Y_{B,k}^{i_H^k} - \omega_k H_{A-B,k}^{i_L^k} Z_{B,k} = H_{A-B,k}^{i_H^k} \left(X_k + \omega_k N_{A,k}^{i_L^k}\right) + N_{B,k}^{i_H^k} \quad (12)$$

The received signal at Bob in (12) is affected by the AGN of Alice forwarded with Bob's AN signal, i.e., the term $H_{A-B,k}^{i_H^k} \omega_k N_{A,k}^{i_L^k}$.

### A. single-link Eavesdropping

When Eve is only eavesdropping on one medium, say medium $i_E = i_H^k$ ($i_E \in \{P,W\}$), her received signal is

$$Y_{E,k}^{i_H^k} = H_{A-E,k}^{i_H^k} \left(X_k + \omega_k \left(H_{A-B,k}^{i_L^k} Z_{B,k} + N_{A,k}^{i_L^k}\right)\right) + N_{E,k}^{i_H^k} \quad (13)$$

If Eve is eavesdropping on medium $i_E = i_L^k$, the eavesdropper receives AGN only since this medium is unused over the $k$-th sub-channel. The rates of the Alice-Bob and Alice-Eve links at sub-channel $k$ are thus given, respectively, by

$$R_{A-B,k} = \mathbb{I}_{i_H^k} \log_2 \left(1 + \frac{|H_{A-B,k}^{i_H^k}|^2 \theta \Gamma_{A-B,k}^{i_H^k}}{|H_{A-B,k}^{i_H^k}|^2 \frac{(1-\theta)\Gamma_{A-B,k}^{i_H^k}}{|H_{A-B,k}^{i_L^k}|^2 \Gamma_{B-A,k}^{i_L^k}+1} + 1}\right) \quad (14)$$
$$+ (1 - \mathbb{I}_{i_H^k}) \log_2 \left(1 + |H_{A-B,k}^{i_H^k}|^2 \Gamma_{A-B,k}^{i_H^k}\right)$$

and

$$R_{A-E,k} = \mathbb{I}_{i_H^k} \log_2 \left(1 + \frac{|H_{A-E,k}^{i_H^k}|^2 \Gamma_{A-B,k}^{i_H^k}}{|H_{A-E,k}^{i_H^k}|^2 (1-\theta) \Gamma_{A-B,k}^{i_H^k} + 1}\right) \quad (15)$$

where $k \in \mathcal{A}$. The ISR of the AN-sharing scheme under single-link eavesdropping is given by

$$R_{A-B}^{\text{sec}} = \left[\sum_{k \in \mathcal{A}} \mathbb{I}_{i_H^k} \log_2 \left(1 + \frac{|H_{A-B,k}^{i_H^k}|^2 \theta \Gamma_{A-B,k}^{i_H^k}}{|H_{A-B,k}^{i_H^k}|^2 \frac{(1-\theta)\Gamma_{A-B,k}^{i_H^k}}{|H_{A-B,k}^{i_L^k}|^2 \Gamma_{B-A,k}^{i_L^k}+1} + 1}\right) \right.$$
$$+ (1 - \mathbb{I}_{i_H^k}) \log_2 \left(1 + |H_{A-B,k}^{i_H^k}|^2 \Gamma_{A-B,k}^{i_H^k}\right)$$
$$\left. - \sum_{k \in \mathcal{A}} \mathbb{I}_{i_H^k} \log_2 \left(1 + \frac{|H_{A-E,k}^{i_H^k}|^2 \theta \Gamma_{A-B,k}^{i_H^k}}{|H_{A-E,k}^{i_H^k}|^2 (1-\theta) \Gamma_{A-B,k}^{i_H^k} + 1}\right)\right]^+ \quad (16)$$

### B. Two-Link Eavesdropping

With two eavesdroppers (i.e., two-link eavesdropping), the rates of the Alice-Bob and Alice-Eve links at sub-channel $k$ are thus given, respectively, by

$$R_{A-B,k} = \log_2 \left(1 + \frac{|H_{A-B,k}^{i_H^k}|^2 \theta \Gamma_{A-B,k}^{i_H^k}}{|H_{A-B,k}^{i_H^k}|^2 \frac{(1-\theta)\Gamma_{A-B,k}^{i_H^k}}{|H_{A-B,k}^{i_L^k}|^2 \Gamma_{B-A,k}^{i_L^k}+1} + 1}\right) \quad (17)$$

and

$$R_{A-E,k} = \log_2 \left(1 + \frac{|H_{A-E,k}^{i_H^k}|^2 \theta \Gamma_{A-B,k}^{i_H^k}}{|H_{A-E,k}^{i_H^k}|^2 (1-\theta) \Gamma_{A-B,k}^{i_H^k} + 1}\right) \quad (18)$$

where $k \in \mathcal{A}$.

Since each Eve is eavesdropping on a medium, the received signal at Eve over medium $i_L^k$ and $i_H^k$, respectively, are given by

$$Y_{E,k}^{i_L^k} = H_{B-E,k}^{i_L^k} Z_{B,k} + N_{E,k}^{i_L^k}$$
$$Y_{E,k}^{i_H^k} = H_{A-E,k}^{i_H^k} \left(X_k + \omega_k \left(H_{A-B,k}^{i_L^k} Z_{B,k} + N_{A,k}^{i_L^k}\right)\right) + N_{E,k}^{i_H^k} \quad (19)$$

Hence, the rate of the Alice-Eve link at sub-channel $k$ is

$$R_{A-E,k} = \log_2 \left(1 + \frac{|H_{A-E,k}^{i_H^k}|^2 \theta \Gamma_{A-B,k}^{i_H^k}}{|H_{A-E,k}^{i_H^k}|^2 (1-\theta) \Gamma_{A-B,k}^{i_H^k} + 1}\right) \quad (20)$$

where $k \in \mathcal{A}$. The ISR of the AN-sharing scheme under the two-link eavesdropping is given by

$$R_{A-B}^{\text{sec}} = \left[\sum_{k \in \mathcal{A}} \log_2 \left(1 + \frac{|H_{A-B,k}^{i_H^k}|^2 \theta \Gamma_{A-B,k}^{i_H^k}}{|H_{A-B,k}^{i_H^k}|^2 \frac{(1-\theta)\Gamma_{A-B,k}^{i_H^k}}{|H_{A-B,k}^{i_L^k}|^2 \Gamma_{B-A,k}^{i_L^k}+1} + 1}\right) \right.$$
$$\left. - \sum_{k \in \mathcal{A}} \log_2 \left(1 + \frac{|H_{A-E,k}^{i_H^k}|^2 \theta \Gamma_{A-E,k}^{i_H^k}}{|H_{A-E,k}^{i_H^k}|^2 (1-\theta) \Gamma_{A-E,k}^{i_H^k} + 1}\right)\right]^+ \quad (21)$$

<u>Remark 1</u>: Over each OFDM sub-channel, the AN-sharing scheme benefits from media diversity by selecting the better link to transmit data. Moreover, there is no rate loss due to transmitting the AN over the other link which was not used in the hybrid parallel PLC/wireless systems with no AN.

From (21), the transmit power level at Bob, included in the SNR level $\Gamma_{B-A,k}^{i_L^k}$, only affects the rate of the Alice-Bob link. Moreover, the ISR in (21) is monotonically increasing with $\Gamma_{B-A,k}^{i_L^k}$. The intuitive explanation of this fact is as follows. Since Bob sends AN signal to Alice and Alice amplifies-and-forwards her received signal, the quality of the forwarded AN signal depends on $P_B$ or, more specifically, on the ratio between the AN power used at Bob and the AGN power at Alice, which is given by $\Gamma_{B-A,k}^{i_L^k}$. If $\Gamma_{B-A,k}^{i_L^k}$ is high, the signal forwarded by Alice is dominated by the AN signal, which is only known at Bob and can be completely canceled prior to information decoding. However, if $\Gamma_{B-A,k}^{i_L^k}$ is low, the signal forwarded by Alice will be dominated by AGN, which is a random signal and unknown to both Bob and Eve. Hence, the noise signal forwarded by Alice will hurt both Bob and Eve and the ISR will decrease accordingly. In addition, the impact of Alice's power level $P_A$ on the ISR in (21) is dependent on Bob's power level $P_B$. If $P_B$ is high, increasing $P_A$ will definitely increase the ISR since the forwarded signal by Alice will be dominated by the AN signal which is known to Bob. Hence, increasing $P_A$ will only hurt Eve. On the other hand, if $P_B$ is low, increasing $P_A$ hurts both Bob and Eve, and it might be the case that increasing $P_A$ hurts Bob more than Eve if the Alice-Bob channel gain is higher than the Alice-Eve channel gain over the lower-CNR medium.

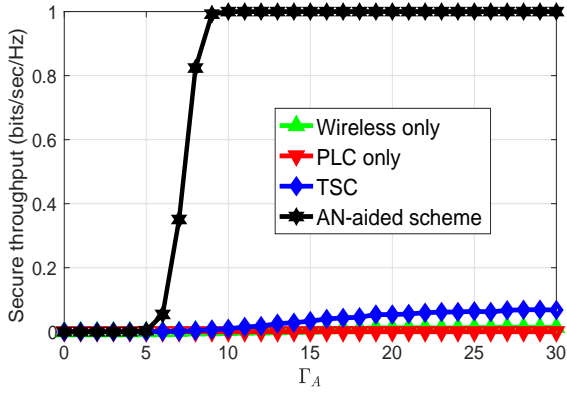

Fig. 2: Secure throughput versus Alice input SNR).

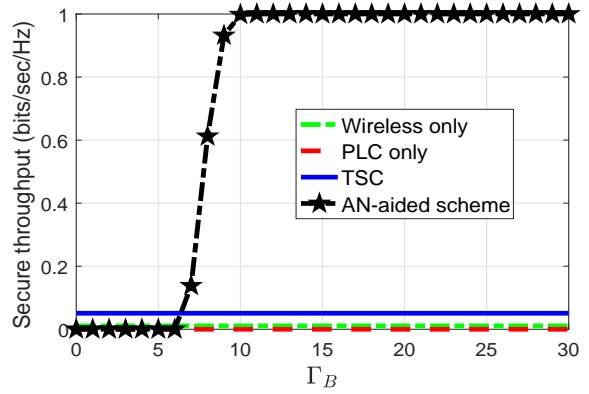

Fig. 3: Secure throughput versus Bob's AN-sharing input SNR.

At infinite $P_A$, (21) reduces to

$$R_{A-B}^{sec} = \left[\sum_{k \in \mathcal{A}} \log_2\left(1 + \frac{\alpha_k \frac{P_B}{|\mathcal{A}|}+1}{\tilde{\theta}}\right) - |\mathcal{A}|\log_2\left(\frac{1}{1-\theta}\right)\right]^+ \quad (22)$$

where $\alpha_k = \frac{|H_{A-B,k}^{i_L^k}|^2}{\kappa_{B,k}^{i_L^k}}$ and $\tilde{\theta} = \frac{(1-\theta)}{\theta}$. It is noteworthy that the ISR at infinite $P_A$ is independent of Eve's CSI (i.e. the rate of the Alice-Eve link will be $|\mathcal{A}|\log_2 \frac{1}{1-\theta}$ which is completely determined by the choice of $\theta$ at Alice). This is an interesting result since Alice can adjust her secrecy rate in each coherence interval to secure her transmissions. That is, in each coherence interval, Alice sets $\mathcal{R}$ to $R_{A-B}^{sec}$.

For fixed-data rate applications where Alice transmits with a target secrecy rate $\mathcal{R}$ or when Alice wishes to send her data with a positive secrecy rate $\mathcal{R}$ by setting $R_{A-B}^{sec} \geq \mathcal{R}$, we will derive a sufficient condition on Bob's transmit power to mitigate the secrecy rate outage events, i.e., the cases $R_{A-B}^{sec} < \mathcal{R}$. The only term that depends on the sub-channel index $k$ in (22) is $\alpha_k$. Since the logarithmic function is monotonically decreasing with its argument, to obtain a lower-bound on the ISR, we can replace all $\alpha_k$ in (22) with the minimum among all $\{\alpha_k\}_{k=1}^{|\mathcal{A}|}$ over all active sub-channels. That is, we set $\alpha_k = \alpha = \min_{\ell \in \mathcal{A}} \alpha_\ell$. Hence, a lower-bound is given by

$$R_{A-B}^{sec} \geq |\mathcal{A}|\left(\log_2\left(1 + \frac{\alpha \frac{P_B}{|\mathcal{A}|}+1}{\tilde{\theta}}\right) - \log_2\left(\frac{1}{1-\theta}\right)\right) \quad (23)$$

To mitigate the secrecy outage events in a given coherence interval or to achieve a positive secrecy rate of $\mathcal{R}$ bits per OFDM block, a sufficient condition is to make the lower-bound on the secrecy rate higher than the target secrecy rate $\mathcal{R}$. Hence, after some manipulations, the condition on Bob's transmit power to mitigate the secrecy outage events is

$$P_B \geq |\mathcal{A}|\frac{\tilde{\theta}\left(2^{\frac{\mathcal{R}}{|\mathcal{A}|}}\left(\frac{1}{1-\theta}\right)-1\right)-1}{\alpha} \quad (24)$$

By setting $\mathcal{R} = 0$ in (24), we find that $P_B > 0$ is a sufficient condition to achieve a non-zero ISR. This further demonstrates the gain of our proposed AN-sharing scheme over the no-AN scenario which may achieve a zero ISR for some channel realizations as argued below (6).

Remark 2: Our proposed AN-sharing scheme cannot be less secure than the no-AN sharing scheme since the latter is a special case of the former by setting $\theta$ to 1. More specifically, substituting with $\theta = 1$ in (21), we obtain the ISR of the no-AN system in (6).

## V. SIMULATION RESULTS

In this section, we investigate the performance of the proposed AN-sharing scheme and compare it with the TSC and single-link schemes. The PLC channel frequency response follows a deterministic channel model obtained from [12] while the wireless channel amplitude gain follows a Rayleigh distribution. Denote the delay spread of the $m_1 - m_2$ link of the PLC medium by $\nu_{m_1-m_2}^P$. We assume that the cyclic prefix (CP) length is $N_{cp} = \nu_{m_1-m_2}^P = \nu_{m_1-m_2}^W = 16$. Unless otherwise stated, we use the system parameters in [8], $\kappa_{m,k}^\ell = \kappa$, $\sigma_{i,m_1-m_2}^2 = 1/(N_{cp}+1)$, $\mathcal{R} = 1$ bits/sec/Hz, $N = 64, \theta = 1/2, \Gamma_A = \frac{P_A}{N\kappa N} = 20$ dB, $\Gamma_B = \frac{P_B}{N\kappa N} = 20$ dB. The simulations are performed under the worst-case scenario assumption of two-link eavesdropping. The secure throughput is measured in bits/sec/Hz which is obtained by dividing the expression in (8) by $(N + N_{cp})$. We quantify the following four important system performance aspects: 1) the secure throughput gain when using a wireless medium in addition to the PLC medium; 2) the secure throughput gain due to data/AN power allocation optimization at Alice; 3) the impact of input SNR at both Alice and Bob on the secure throughput; and 4) the secure throughput gain of our proposed AN-sharing scheme relative to the no-AN scenario.

In Fig. 2, we compare the secure throughput of the hybrid parallel system, both with and without AN, with the PLC-only and wireless-only systems. As Alice's transmit power level increases (i.e., as $\Gamma_A$ increases), the secure throughput is non-decreasing for all systems. Our proposed scheme achieves the upper bound performance where the achievable secure throughput is equal to the target secrecy rate $\mathcal{R} = 1$ bits/sec/Hz. The wireless-only system outperforms the PLC-only system since the wireless link follows a statistical distribution and therefore it has higher diversity than the PLC channel which is assumed to have a deterministic nature. TSC outperforms the PLC-only and wireless-only systems due to the diversity introduced by combining the two media. The proposed AN-sharing scheme significantly outperforms the TSC scheme since in the AN-sharing case the eavesdropper suffers heavily from the injected AN. In the TSC scheme, Bob and Eve sub-channels have similar average power gains and Alice selects the sub-channels to Bob that have higher CNRs. In contrast, the AN-sharing scheme degrades Eve's SNR and offers SNR gains for Bob over Eve. The only price paid in our proposed scheme is that Bob will use its power to share the AN with Alice.

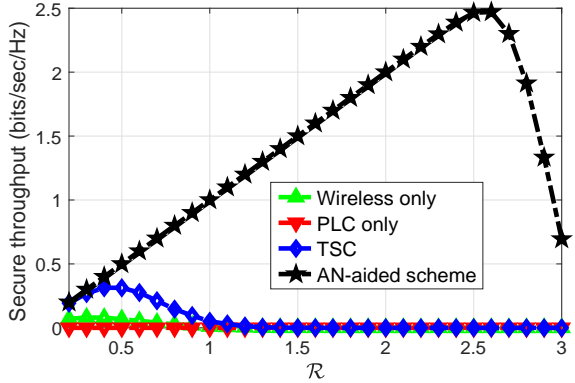

Fig. 4: Secure throughput with respect to the target (required) secrecy rate $\mathcal{R}$.

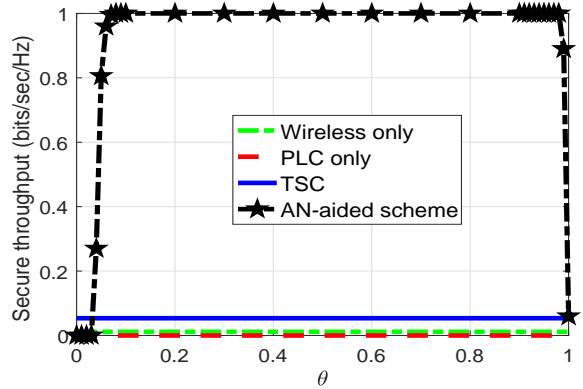

Fig. 5: Secure throughput with respect to Bob's AN sharing input SNR.

Fig. 3 shows that the secure throughput for the AN-sharing scheme increases with $\Gamma_B$, which corroborates our discussions in Section IV. At low $\Gamma_B$ levels, Alice's AGN, which is forwarded by Alice with the AN signal, becomes comparable to the AN power and degrades Bob's ability to cancel the AN. That is, the forwarded AN signal by Alice is dominated by the AGN which is random at both Bob and Eve and degrades the received signals at both of them. In contrast, as $\Gamma_B$ increases, Alice's AGN transmitted with the AN becomes negligible. Hence, Bob cancels the AN effectively.

Fig. 4 depicts the secure throughput versus the target secrecy rate $\mathcal{R}$. As expected, for all scenarios, the secure throughput increases linearly with $\mathcal{R}$ until the optimal $\mathcal{R}$ is reached, then it starts decreasing. The optimal value for $\mathcal{R}$ of our proposed AN-aided scheme is around $2.5$ bits/sec/Hz. This shows a significant secure throughput gain compared to TSC which has a maximum secure throughput around $\mathcal{R} = 0.5$ bits/sec/Hz.

Fig. 5 shows the secure throughput as a function of $0 \leq \theta \leq 1$. The figure demonstrates that our AN-aided scheme will achieve the required secure throughput as long as the transmission power is balanced between data and AN transmissions. The case of $\theta = 1$ corresponds to allocating all transmit power to data and zero power to AN which is the TSC case. This further shows the throughput gain of our scheme relative to TSC.

## VI. CONCLUSIONS

To enhance the security of OFDM-based hybrid parallel PLC/wireless systems, we proposed an AN-aided scheme which exploits the decoupled nature of the PLC and wireless media. We showed that the ISR of our proposed scheme, at infinite Alice transmit power level, is independent of the eavesdropper's CSI. Moreover, our proposed scheme can achieve any target secrecy rate by adjusting the transmit power levels at both Alice and Bob. Our numerical results quantified the appreciable secure throughput gains of our proposed AN-sharing scheme compared to the no-AN scenario for different operating conditions.

ACKNOWLEDGEMENT

We thank Prof. Lutz Lampe of University of British Columbia (UBC) for many stimulating discussions and for his comments on earlier versions of this paper.

REFERENCES

[1] M. Sayed, T. A. Tsiftsis, and N. Al-Dhahir, "On the diversity of hybrid narrowband-PLC/wireless communications for smart grids," *IEEE Trans. Wireless Commun.*, vol. 16, no. 7, pp. 4344–4360, July 2017.
[2] S. W. Lai and G. G. Messier, "Using the wireless and PLC channels for diversity," *IEEE Trans. Commun.*, vol. 60, no. 12, pp. 3865–3875, Dec. 2012.
[3] S. W. Lai, N. Shabehpour, G. G. Messier, and L. Lampe, "Performance of wireless/power line media diversity in the office environment," in *Proc. IEEE Globecom*, Dec. 2014, pp. 2972–2976.
[4] H. Su, M. Qiu, and H. Wang, "Secure wireless communication system for smart grid with rechargeable electric vehicles," *IEEE Commun. Mag.*, vol. 50, no. 8, pp. 62–68, 2012.
[5] N. Saxena, S. Grijalva, V. Chukwuka, and A. V. Vasilakos, "Network security and privacy challenges in smart vehicle-to-grid," *IEEE Wireless Commun.*, vol. 24, no. 4, pp. 88–98, 2017.
[6] A. D. Wyner, "The wire-tap channel," *Bell Syst. Tech. J.*, vol. 54, no. 8, pp. 1355–1387, 1975.
[7] A. Pittolo and A. Tonello, "Physical layer security in PLC networks: Achievable secrecy rate and channel effects," in *IEEE ISPLC*, 2013, pp. 273–278.
[8] Y. Zhuang and L. Lampe, "Physical layer security in MIMO power line communication networks," in *IEEE ISPLC*, Mar. 2014, pp. 272–277.
[9] A. Pittolo and A. Tonello, "Physical layer security in power line communication networks: an emerging scenario, other than wireless," *Commun., IET*, vol. 8, no. 8, pp. 1239–1247, May 2014.
[10] A. Mukherjee, S. Fakoorian, J. Huang, and A. Swindlehurst, "Principles of physical layer security in multiuser wireless networks: A survey," *IEEE Commun. Surveys Tuts.*, vol. 16, no. 3, pp. 1550–1573, 2014.
[11] W. Rhee and J. M. Cioffi, "Increase in capacity of multiuser OFDM system using dynamic subchannel allocation," in *Proc. IEEE VTC*, vol. 2, 2000, pp. 1085–1089.
[12] A. M. Tonello and T. Zheng, "Bottom-up transfer function generator for broadband plc statistical channel modeling," in *Power Line Communications and Its Applications, 2009. ISPLC 2009. IEEE International Symposium on*. IEEE, 2009, pp. 7–12.